\documentclass[conference,letterpaper]{IEEEtran}
\usepackage[utf8]{inputenc} 
\usepackage[T1]{fontenc}
\usepackage{url}
\usepackage{ifthen}
\usepackage{cite}
\usepackage[cmex10]{amsmath}
\usepackage{amssymb}
\usepackage{quantikz}
\usepackage{subfigure}
\usepackage{subcaption}
\usepackage{graphicx}
\usepackage{colortbl}
\usepackage{tabularx}
\usepackage{multirow}

\newcolumntype{Y}{>{\centering\arraybackslash}X}
\newcolumntype{C}{>{$}c<{$}}
\newcolumntype{L}{>{$}l<{$}}
\newcolumntype{s}{>{\columncolor{yellow}}p{4.5cm}}
\renewcommand{\arraystretch}{1.3}
\newtheorem{theorem}{Theorem}
\newtheorem{definition}{Definition}

\begin{document}
\title{Fault-tolerance of the [[8,1,3]] non-CSS code} 

\author{\IEEEauthorblockN{Pranav Maheshwari\IEEEauthorrefmark{1}
and
Ankur Raina\IEEEauthorrefmark{2}}
\IEEEauthorrefmark{1}Department of Physics, Indian Institute of Science Education and Research Pune, India \\
\IEEEauthorrefmark{2}Department of EECS, Indian Institute of Science Education and Research Bhopal, India\\
Email: \IEEEauthorrefmark{1}pranav.maheshwari@students.iiserpune.ac.in,
\IEEEauthorrefmark{2}ankur@iiserb.ac.in}

\maketitle

\begin{abstract}
   We present a fault-tolerant $[[8,1,3]]$ non-CSS quantum error correcting code and study its logical error rates.  
   We choose the unitary encoding procedure for stabilizer codes given by Gottesman and modify it to suit the setting of a class of non-CSS codes.
   Considering two types of noise models for this study, namely the depolarising noise and anisotropic noise, to depict the logical error rates obtained in decoding, we adopt the procedure of the bare ancilla method presented by Brown \emph{et al.} to reorder the measurement sequence in the syndrome extraction step and upgrade it to obtain higher pseudo-thresholds and lower leading order terms of logical error rates.
   
\end{abstract}

\section{Introduction}
% \paragraph{Paragraph 1: General utility of QECC}
While quantum computing boasts immense power, it contends with a relentless adversary: environmental noise that disrupts quantum systems. 
This challenge is particularly pronounced in NISQ (Noisy Intermediate-Scale Quantum) computers, which must grapple with error correction to conduct meaningful computations amidst this noise.
Quantum Error Correcting Codes (QECC) are specialized encoding schemes that distribute quantum information redundantly across multiple qubits in such a way that errors can be detected and corrected.
QECC are organized into distinct families; a well-studied family is of Calderbank-Shor-Steane (CSS) Codes which treats $X$ and $Z$ Pauli errors separately\cite{CSS_shor}\cite{CSS_Steane}.

% \paragraph{Paragraph 2: Fault tolerance introduction and Bare ancilla method of Brown et al. }

Integrating error correction introduces errors due to noisy gates, thereby underscoring the significance of fault tolerance.
Fault tolerance ensures that the system remains resilient against these newly induced errors, providing an additional layer of protection to maintain the integrity of the quantum computation.
The Threshold theorem in quantum computing asserts that if the error rate in quantum computation is below a certain threshold, then it is possible to perform arbitrarily long quantum computations reliably by using QECC \cite{aharonov}\cite{nielsen_chuang_2010}.
Many fault-tolerant schemes have been proposed that require two or more ancillary qubits \cite{ft_shor}\cite{ft_steane}\cite{ft_2qubits}.
Brown \emph{et. al} proposed a fault-tolerant seven-qubit code that requires a single ancilla to correct any error propagated from the syndrome qubit\cite{bare_ancilla}.

% \paragraph{Paragraph 3: Our approach compared to that of Brown et al and what is novel, what are the main contributions of the paper? }

We discovered an eight-qubit non-CSS code utilizing cluster states measurements \cite{shaw} and molded it to obtain a bare ancillary fault-tolerant code.
We upgraded the bare ancillary method in \cite{bare_ancilla} to correct all two-qubit errors and discarded the limitations poised by hook errors.
We reveal a unitary Encoder for non-CSS codes and simulate the implementation of any quantum error correction code using IBM's qiskit \cite{Qiskit}.
We sketch the procedure to obtain pseudo-threshold values and the leading order of corrected error rates.

This paper studies the fault-tolerant properties of the eight-qubit non-CSS code under the standard depolarising noise model and the anisotropic noise model.
We test a physically suitable error correction procedure against the traditional simulation scheme and obtain pseudo-thresholds and the leading order term in error rates after quantum error correction (QEC).

% \paragraph{Organization of the paper}
The paper is organized as follows: Section~\ref{sec:QEC} introduces quantum error correction and implementation procedure.
Section~\ref{sec:encoder} describes the unitary encoding scheme and its enhanced version that is suitable for non-CSS codes. 
Section~\ref{sec:single_qubit_ft} introduces the eight-qubit code and details the single-qubit fault tolerance method.
It explains how the code can bypass any propagated error by reordering stabilizer measurements and the corresponding Encoder and Detector circuits.
Noise Models of our interest are outlined in Section~\ref{sec:noise_models} followed by simulation schemes in Section~\ref{sec:simulation}. We present one complete cycle of bare ancillary error correction and its modification to obtain pseudo-threshold values.
In Section~\ref{sec:results}, we discuss the code performance and the reason why the standard depolarising model shows more optimistic error rate values. We also comment on practical QEC procedures and why modifications are required in simulations to obtain benchmarks.

\subsection{Notation}
We list the notations used in this work:\\

\begin{tabular}{@{}ll@{}}
$n$ &  Number of physical qubits.\\
$k$ & Number of logical qubits.\\
$d$ & Minimum Hamming distance of the code.\\
$\mathcal{S}$ & Stabilizer subgroup.\\
$g$ & Generators of the stabilizer group.\\
$M$ & Standard form generators\\
$\mathcal{C}$ & Code or Codespace. \\
$X_i,Y_j,Z_k$ & Pauli matrices on  qubits $i,j,k$ respectively.\\
$H$ & Check matrix of the code.\\
$I$ & Identity operator \\
$\mathbf{H}$ & Hadamard gate.\\
$\mathbf{S}$ & Phase gate. \\
$\ket{\psi_i}$ &  $k$ qubit initial state\\
$\ket{\psi_e}$ &  $n$ qubit encoded state\\
$\ket{\psi_d}$ &  $n$ qubit decoded state\\
$\mathbb{F}_2$ & Binary field $\{0,1\}$
\end{tabular}

\section{Quantum error correction}
\label{sec:QEC}
% \textcolor{red}{Add more content. This section is too small. Start section 2 on page 2. Expand the section 1 to one full page. }
First of the QECC were discovered independently by Shor \cite{shor_code} and Steane \cite{steane_code}, followed by extensive work to create quantum codes from classical codes.
This section provides insights into classical coding theory and the structure of QECC.

\subsection{Preliminaries}
In classical error correction, a \textit{linear code, $\mathcal{C}$}, encodes $k$ bits of information into $n$ bits and is denoted as an $[n,k]$ code.
The code is specified by a matrix of size $n \times k$ called the \textit{generator matrix}, $G$, whose entries are elements of $\mathbb{F}_2$.
To encode all messages uniquely, we require that columns of $G$ be linearly dependent.

A codespace of an $[n,k]$ code is defined to consist of all vectors of length $n$, $x \in \mathbb{F}_2^n$, such that $Hx = 0$ where $H$ is a matrix of size $n-k \times n$, known as \textit{Parity Check Matrix}.
We can also say that codespace is the \textit{kernel space} of $H$. 
By \textit{Rank-Nullity Theorem}, $H$ should have linearly independent rows.
The matrices $G$ and $H$ are related as \textbf{$HG=0$}.

\textit{Hamming Distance, d(x,y)} between two codewords $x$ and $y$ is defined to be the number of places at which $x$ and $y$ differ.
The 
\textit{weight} of a codeword (any vector in codespace) is defined to be the distance from the string of all zeroes, i.e.,
\begin{equation}
   \text{wt}(x)= d(x,0_n) \quad \forall x \in \mathcal{C}. 
\end{equation}
\textit{Distance of a code}, $d$ is the minimum distance between any two codewords.
\begin{equation}
    d(\mathcal{C})=\min \{ d(x,y) \quad | \quad x,y \in \mathcal{C} \;\; \text{and} \;\; x \neq y\},
\end{equation}
which implies $d(\mathcal{C})= \min \{ \text{wt} (x) | x \in \mathcal{C} \}$.
This defines an $[n,k,d]$ code.

Consider a codeword $y=Gx$, but an error is induced such that $y'=y+e$.
Since $Hy=0$, we get $Hy'=He=s$, called the error syndrome.
If for all possible errors $e_j$, we know the corresponding $s_j$, we can match the error syndrome obtained to identify the error and perform error correction accordingly.
For a code to correct errors up to weight $t$, the minimum distance of the code should be $2t+1$ so that $y'$ can be corrected to $y$ satisfying $d(y,y') \leq t$.

In quantum computation, $n$ length vectors are replaced by $n$ qubit state vectors, and operators replace the matrices.
Akin to the parity check matrix, $H$, we create a $n-k \times 2n$ check matrix with independent rows.
These rows correspond to $n-k$ independent operators acting on $n$ qubits, called the \textit{Stabilizer generators}.
The span of these operators generates a group called the \textit{Stabilizer subgroup}, $\mathcal{S}$.
QECC harnessing stabilizers are called \textit{Stabilizer Codes}, and we refer to the corresponding error correction mechanism as \textit{Stabilizer Formalism} \cite{gottesman_thesis}.

A single qubit can exist in a $2^1$ dimensional complex vector space, denoted by $\mathcal{V}_1$.
Correspondingly, the operators acting on a single qubit state can be expressed as a linear combination of \textit{Pauli matrices} forming a basis group, $\mathcal{G}_1$.
For a single qubit, we can define \textit{Pauli Basis Group} as:
\begin{equation} 
    \mathcal{G}_1 \equiv \{\pm I, \pm iI, \pm X, \pm iX, \pm Y, \pm iY, \pm Z, \pm iZ\}.
\end{equation}
This group can be generated by operators $iI, X, Z$ represented as $\mathcal{G}_1 = \langle iI, X, Z\rangle$.
For $n$ qubits, Pauli group ($\mathcal{G}_n$) contains $n$-fold tensor products of Pauli matrices with multiplicative factors $\{\pm 1, \pm i\}$.

Tensor products of Pauli matrices follow the following two properties:
\begin{enumerate}
    \item For any two elements $g_1, g_2 \in \mathcal{G}_n$, either they commute or anti-commute, i.e., either $[g_1, g_2] = 0$ or $\{g_1, g_2\} = 0$. 
    \item Each element is Hermitian or Anti-Hermitian i.e., $\forall g \in \mathcal{G}_n, \; g^{\dagger} = \pm g$. Also, $g^2 = \pm I$.
\end{enumerate}

\begin{theorem}
    Let $\mathcal{S} = \langle g_1, \dots, g_{n-k}\rangle$ be generated by $n-k$ independent and commuting elements from $\mathcal{G}_n$, such that $-I \notin \mathcal{S}$, then $\mathcal{V}_{\mathcal{S}}$ is a $2^k$ dimensional vector space.
\end{theorem}

This $\mathcal{V}_{\mathcal{S}}$ serves as the codespace for an $[[n,k,d]]$ quantum code containing $k$ qubits information in $n$ qubits.
All the stabilizers commute with each other and all the codewords are in +1 eigenspace of the stabilizers.

\begin{definition}
    Normalizer of a subgroup $\mathcal{S}$ in group $\mathcal{G}_n$ is the the set of elements $\mathcal{N}$ given by
    \begin{equation}
        \mathcal{N}_{\mathcal{G}_n}(\mathcal{S}) = \{g \in \mathcal{G}_n \quad | \quad g \mathcal{S} g^{-1} = \mathcal{S}\}.
    \end{equation}

\end{definition}

Thus, the normalizer subgroup contains all elements that commute with the stabilizer subgroup.
The elements in $\mathcal{N}_{\mathcal{G}_n}(\mathcal{S})$ are also called logical operators.
For an $[[n,k]]$ code, there exists $n-k$ stabilizer generators and $2k$ normalizer generators.
These generators correspond to the encoded $\mathcal{G}_k$ group and are referred as $\bar{X}$ and $\bar{Z}$ operators.

\begin{theorem}
    Let $\mathcal{S}$ be the stabilizer subgroup for a code $\mathcal{C}$. Suppose $\{E_j\}$ is a set of operators in $\mathcal{G}_n$ such that ${E_j}^{\dagger}E_k \notin \mathcal{N}_{\mathcal{G}_n}(\mathcal{S})\setminus\mathcal{S} \; \forall j,k$. Then $\{E_j\}$ is a correctable set of errors for code $\mathcal{C}$.
\end{theorem}

There exists $n-k$ independent error operators, each of which  anticommutes to one of the stabilizer and commutes to all others.
The check matrix provides a non-zero value to syndrome vector of an error corresponding to the row which anticommutes with the error.

$\mathcal{G}_n$ group has $2n$ generators.
If we replace each by its encoded version i.e., $P \rightarrow \bar{P}$ where $P$ is a generator of $\mathcal{G}_n$, Fig.~\ref{fig:stab_alt_pic} provides a simple visualization to Stabilizers ($I$), Normalizers ($\mathcal{G}_k$), and Errors ($\mathcal{E}$).
\begin{figure}[htbp]
    \centering
    \begin{align*}
        \overbrace{\textcolor{blue}{\bar{Z}_1, \dots, \bar{Z}_{n-k}}}^{I} \;\;  \textcolor{cyan}{\bar{Z}_{n-k+1}, \dots, \bar{Z}_{n}}\\
        \underbrace{\textcolor{red}{\bar{X}_1, \dots, \bar{X}_{n-k}}}_{\mathcal{E}} \;\; \underbrace{\textcolor{cyan}{\bar{X}_{n-k+1}, \dots, \bar{X}_{n}}}_{\mathcal{G}_k}
    \end{align*}
    \caption{Visualization of encoded operators in $\mathcal{G}_n$ corresponding to stabilizer formalism.}
    \label{fig:stab_alt_pic}
\end{figure}

A deeper understanding of the stabilizer formalism could be gained from \cite{nielsen_chuang_2010}.
A mathematically extensive explanation is given in \cite{gaitan}.

\subsection{Error Correction Protocol}
\begin{figure}[htbp] 
    \centering
    % \tikzset{noisy/.style={starburst,fill=yellow,draw=red,line width=0.5pt,inner xsep=-4pt,inner ysep=-2pt}}
    \resizebox{255pt}{40pt}{
    \begin{quantikz}[wire types = {q,q,n}]
        \lstick{$\ket{0}$} & \qwbundle{n-k} & \gate[2]{\text{Encoder}} & \gate[3]{\text{Detector}} \gategroup[3,steps=4,style={dashed,rounded corners,fill=blue!20, inner xsep=1pt},background,label style={label position=below,anchor=north,yshift=-0.2cm}]{{Decoder}}
        && \gate[2]{\text{Corrector}} & \gate[2]{\text{(Encoder)}^{\dagger}} & \rstick[2]{$\ket{\psi_d}$}\\
        \lstick{$\ket{\psi_i}$} & \qwbundle{k} &&&&&& \\
        &&\lstick{$\ket{+}$}& \setwiretype{q} & \setwiretype{c} & \ctrl[vertical wire=c]{-1} & \wireoverride{n} & \wireoverride{n} & \wireoverride{n}
    \end{quantikz}
    }
    \caption{Block diagram description of an $[[n,k]]$ quantum error correcting code.}
    \label{fig:error_correction_picture}
\end{figure}
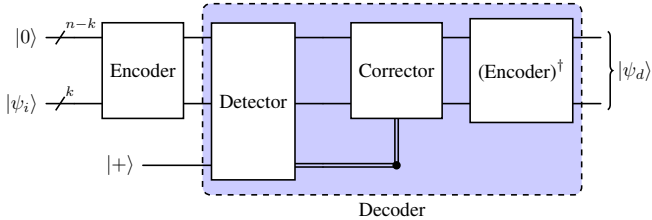

Given the stabilizer generators, $\{g_1, \dots, g_{n-k}\},$ of a  QECC, we encode $k$ logical qubit state information, $\ket{\psi_{i}}$, into an $n$ physical qubit state $\ket{\psi_e}$ that belongs to the codespace $\mathcal{C}$. Such a code can tolerate certain errors during computation.
The entire process of QEC can be summarized as follows:
\begin{enumerate}
    \item Initialize an $n$ qubit state, $\ket{0}^{\otimes n-k} \ket{\psi_{i}}$ where $\ket{\psi_{i}}$ is the $k$ qubit state to be encoded.
    \item Apply a unitary quantum circuit called \textit{Encoder} which maps the above $n$ qubit state to the simultaneous +1 eigenspace of all stabilizer generators, giving the $\ket{\psi_{e}}$ state.
    \item \label{step: s_i_meas} Take an ancilla qubit in $\ket{+}$ state, and apply controlled $g_1$ operation on the encoded state, controlled on the ancilla.
    \item Measure the ancilla qubit in $X$ basis and store the outcome. Reset the ancilla to $\ket{+}$ state and repeat step~\ref{step: s_i_meas} for all $g_i$ where $i \in \{2, \dots, n-k\}$.
    \item Combine the stored outcomes in order and call it the \textit{syndrome}. Feed the syndrome to a corrector function, which applies the appropriate recovery operation.
    \item Since the action of the Encoder is a unitary operation, we can apply inverse of each operator in Encoder, in reverse order, to obtain the decoded state, $\ket{\psi_{d}}$.
\end{enumerate}
A schematic of the steps involved in QEC is presented in Fig.~\ref{fig:error_correction_picture}.
The details of constructing the Encoder is given in Section~\ref{sec:encoder}.
Circuits specific to [[8,1,3]] code are given in Section~\ref{sec:circuits}.
    
\section{Encoder Design}\label{sec:encoder}
Encoding requires embedding a $2^k$ dimensional subspace inside a $2^n$ dimensional Hilbert space $\mathbb{C}^{2^n}$ of $n$ qubits. Step 3 and Step 4, given in Section \ref{sec:QEC}, aim at applying projection operators that take an $n$-qubit state to the simultaneous +1 eigenspace of the stabilizer generators, commonly called the codespace, $\mathcal{C}$.
The Detector and Corrector combined act as an Encoder, which we call the \textit{measurement based Encoder}.
Since it involves non-unitary measurement operations, we cannot use its inverse to decode a codeword state back to $\ket{\psi_{i}}$.
An efficient encoding technique is presented in \cite{gottesman_encoder} which shuffles the stabilizer generators in a certain form so as to obtain an encoding unitary circuit as described next.

\subsection{Efficient encoding mechanism}
The check matrix, $H$ corresponding to the stabilizer generators contains two blocks $H^x$ and $H^z$ corresponding to qubit location of $X$ and $Z$ operators in the stabilizer generators as given in Equation~\ref{eq:check_matrix}:
\begin{equation}\label{eq:check_matrix}
    \mathcal{S} \longrightarrow [H^x_{n-k \times n} | H^z_{n-k \times n}].
\end{equation}
Adding any two rows in $H$ is equivalent to multiplying the respective stabilizer generators, and swapping two columns is equal to relabeling the qubits.
Following these two actions, we can translate any check matrix to the standard form \cite{nielsen_chuang_2010} given in Equation ~\ref{eq:standard_form}, where $r$ is the rank of $H^x$ block and $l=n-k-r$:
\begin{gather}\label{eq:standard_form}
    \left[
    \begin{array}{ccc|ccc}
        I_{r \times r} & A_{r \times l} & A'_{r \times k} & B_{r \times r} & 0_{r \times l} & C_{r \times k}\\
        0_{l \times r} & 0_{l \times l}   & 0_{l \times k} & D_{l \times r} & I_{l \times l} & E_{l \times k}
    \end{array}
    \right].
\end{gather}
We can also obtain logical operators ($\bar{X}$, $\bar{Z}$) from the standard form using Equation~\ref{eq:logical_operators}:
\begin{gather}\label{eq:logical_operators}
    \left[
    \begin{array}{c}
         \bar{X}  \\
         \bar{Z} 
    \end{array}
    \right]
    =
    \left[
    \begin{array}{ccc|ccc}
        0_{k \times r} & E^T_{k \times l} & I_{k \times k} & C^T_{k \times r} & 0_{k \times l} & 0_{k \times k}\\
        0_{k \times r} & 0_{k \times l}   & 0_{k \times k} & A'^T_{k \times r} & 0_{k \times l} & I_{k \times k}
    \end{array}
    \right].
\end{gather}

\begin{figure}[htbp]
    \centering
    \begin{quantikz}
        \lstick[2]{$\ket{0}^{\otimes r}$} & \gate{\mathbf{H}} & \ctrl{1} && \gate{M_2} & \rstick[4]{$\ket{\psi_e}$}\\
        & & \gate[3]{M_1} & \gate{\mathbf{H}} & \ctrl{1}\wire[u]{q} &\\
        \lstick{$\ket{0}^{\otimes l}$} & \gate{\bar{X}} &&& \gate[2]{M_2} &\\
        \lstick{$\ket{\psi_i}\;\;$} & \ctrl{-1} &&&&\\
    \end{quantikz}
    \caption{An schematic of the efficient encoder example for 2 stabilizers $(M_1, M_2)$ in standard form.}
    \label{fig:gottesman_encoder}
\end{figure}
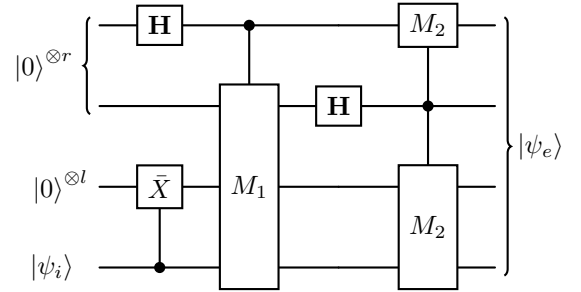

We can retrieve a new set of stabilizer generators, $M_1,\dots,M_{n-k}$, from the standard form.
To create the corresponding Encoder circuit, we initialize $n-k$ ancillae in $\ket{0}$ state and $k$ information qubits in $\ket{\psi_i}$ state.
The action of Encoder gives the encoded state $\ket{\psi_e}$ equivalent to 
\begin{equation}
    \ket{\psi_e} = \alpha \times \prod\limits_{i=1}^{n-k}(I+M_i)\ket{0}^{\otimes n-k}\ket{\psi_i},
\end{equation}
where $\alpha$ is the normalization factor.
Applying Hadamard gate $\mathbf{H}$ on first ancillary qubit followed by controlled-$M_1$ operation, with control on teh ancilla, gives the state $(I+M_1)\ket{0}^{\otimes n-k}\ket{\psi_i}$.
Proceeding similarly for all ancilla up to $r$ gives the $\ket{\psi_i}$ state.
To accommodate logical operations within the Encoder, we first need to apply controlled-$\bar{X}$ operations as depicted in Fig.~\ref{fig:gottesman_encoder} showing an illustration of the Encoder \cite{gottesman_encoder}.

A modification given in \cite{grassl_encoder} points to errors in \cite{gottesman_thesis} and provides suggestions on removing non-essential gates.
In the next section, we adapt the mechanism to suit for encoding non-CSS codes.

\subsection{Enhanced Encoder for non-CSS Codes}
The efficient encoding mechanism in \cite{gottesman_encoder} defines $Y := XZ$ and omits any associated phase. \cite{gaitan}
We notice that $Y^2 = I$ but $(XZ)^2 = -I$, which might change the syndrome values and the encoded state.
Thus, the above method suits perfectly to CSS codes but fails with certain non-CSS codes.

To incorporate the required sign, we add another column, \textit{phase column}, in the check matrix, $H$.
Corresponding to $\{+, i, -, -i\}$, phase of the stabilizer, we add values $\{0,1,2,3\}$ respectively in the last column.
Similar row operations (modulo 4 for phase columns) are applicable leading to the standard phase column of the standard matrix.

Each Pauli $Y$ in the standard form stabilizers devours a multiple of $i$ from the phase column.
We can then obtain the standard stabilizer generators and corresponding phases.
While creating circuit, we add an $\mathbf{S}$ gate after $\mathbf{H}$ gate to encode $(I+Y)\ket{0}$.
And to account for stabilizer phase, add an $\mathbf{S}, Z$ or $\mathbf{S}Z$ gate respectively for $i, -1$, and $-i$ in phase column.
We check if the state has been mapped to +1 eigenspace of all stabilizer generators.
Through additional $\mathbf{S}$ and $Z$ operators, we can correct any phase difference in the encoded state leading to unexpected syndrome values.

\section{Single Qubit Fault Tolerance}\label{sec:single_qubit_ft}
Perfect codes such as the [[5,1,3]] code saturates the quantum Hamming bound, i.e., utilizes all possible syndrome values to detect errors correctable by the code.
For an $[[n,k,d]]$ code, we expect $2^{n-k}$ distinct syndrome values.
A non-degenerate code of distance, $d=3$, requires $3\times$ $k\choose 1$ distinct syndrome values for error correction.
The difference of total possible values and the required syndrome values are the spare syndromes which could be leveraged to detect faults that occur during the error correction protocol.
Utilizing those values of syndrome measurements sets the footing of single-qubit fault tolerance method that requires a single ancillary qubit to correct errors due to fault gates which otherwise would lead to higher weight errors or logical errors.

Following the syndrome detection method as outlined above, while obtaining syndrome value corresponding to a stabilizer, say, $M=X_1Y_2Z_3$, a $P \otimes X$ error on the ancilla qubit after the three gates propagates to the following errors: $P_1Y_2Z_3$, $P_2Z_3$, and $P_3$ respectively as shown in figure~\ref{fig:bare_normal_detector}.
Here $P_i \in \{I_i, X_i, Y_i, Z_i\}$.
For fault tolerance, we ensure that a single fault must lead only to an error correctable by the code.
Thus, in addition to errors correctable by the distance property of the code, we only need to detect the data qubit errors arising due to faulty gates.
This requires each propagated error to have syndrome values distinct from the correctable errors as well as other propagated errors.
  
\subsection{Reordering Stabilizers}
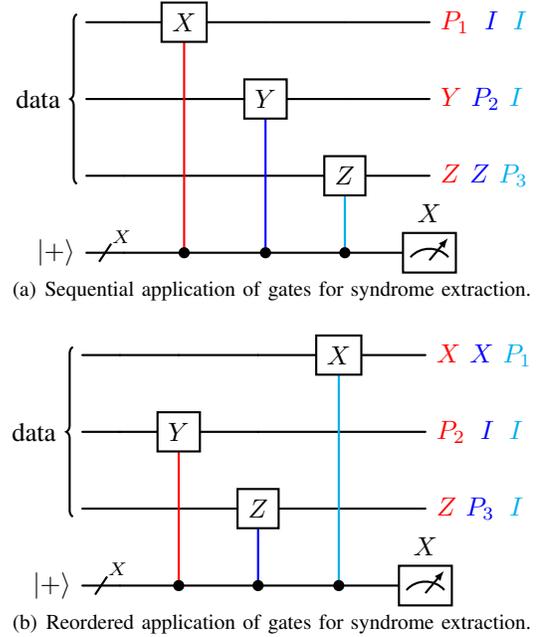
\begin{figure}[htbp]
    \centering
    % \centerline{
        \subfigure[Sequential application of gates for syndrome extraction.]{
            % \resizebox{220pt}{65pt}{
            \begin{quantikz}
                \lstick[3]{data} && \gate{X} &&& \rstick{\textcolor{red}{$P_1$} \textcolor{blue}{$\:I$} \textcolor{cyan}{$\:I$}}\\
                &&& \gate{Y} && \rstick{\textcolor{red}{$Y$} \textcolor{blue}{$P_2$} \textcolor{cyan}{$I$}}\\
                &&&& \gate{Z} & \rstick{\textcolor{red}{$Z$} \textcolor{blue}{$Z$} \textcolor{cyan}{$P_3$}}\\
                \lstick{$\ket{+}$} & \qwbundle{X} & \ctrl[wire style={red}]{-3} & \ctrl[wire style={blue}]{-2} & \ctrl[wire style={cyan}]{-1} & \meter{X}
            \end{quantikz}%}
            \label{fig:bare_normal_detector}
        }
        \subfigure[Reordered application of gates for syndrome extraction.]{    
            % \resizebox{220pt}{65pt}{
            \begin{quantikz}
                \lstick[3]{data} &&&& \gate{X} & \rstick{\textcolor{red}{$X$} \textcolor{blue}{$X$} \textcolor{cyan}{$P_1$}}\\
                && \gate{Y} &&& \rstick{\textcolor{red}{$P_2$} \textcolor{blue}{$\:I$} \textcolor{cyan}{$\:I$}}\\
                &&& \gate{Z} && \rstick{\textcolor{red}{$Z$} \textcolor{blue}{$P_3$} \textcolor{cyan}{$\:I$}}\\
                \lstick{$\ket{+}$} & \qwbundle{X} & \ctrl[wire style={red}]{-2} & \ctrl[wire style={blue}]{-1} & \ctrl[wire style={cyan}]{-3} & \meter{X}
            \end{quantikz}%}
            \label{fig:bare_reordered_detector}
        }
    % }
    \caption{Propagation of error on data qubits when two qubit gates are followed by a $P \otimes X$ error, where $P \in \{X,Y,Z\}$. Errors of the kind $P \otimes Z$ lead only to $P$ error on respective data qubit as $Z$ error on ancilla qubit does not propagate to data qubits. Each color denotes the gate that gets faulty and the corresponding error on data qubits.}
    \label{fig:bare_reordering}
\end{figure}

In certain cases, the syndrome values corresponding to faults on one stabilizer may match the syndrome of fault on another stabilizer or a correctable error.
We use the reordering trick to eliminate such similarities. 
We notice how the propagated errors change if we reorder $M$ to $Y_2Z_3X_1$ as shown in figure~\ref{fig:bare_reordered_detector}.
We shall reorder the gates such that all propagated errors have distinct syndromes.
A similar technique is used in \cite{bare_ancilla} to achieve fault tolerance under anisotropic noise model.

Consider a stabilizer of form $S_iS_jS_kS_l$, if a $P \otimes X$ error occurs on gate corresponding to $S_i$, the propagated data error will be $P_iS_jS_kS_l \equiv P'_i$ which is a single qubit pauli error on $i^{\mathrm{th}}$ qubit.
Similarly error on controlled $S_j$ gate leads to $P_jS_kS_l \equiv S_iP'_j$ error where $P' = (PS)$ is another Pauli operator.
Considering a code that can correct any single qubit error, we are concerned only about propagated errors of weight two or more.
We observe that a weight four stabilizer of form $S_iS_jS_kS_l$ can only lead to $S_iP'_j$ and $P'_kS_l$ kind of higher weight errors while a weight six stabilizer $S_iS_jS_kS_lS_mS_n$ can only lead to $S_iP'_j$, $S_iS_jP'_k$, $P'_lS_mS_n$, and $P'_mS_n$ kind of higher weight errors where $S, P, P' \in \{I, X, Y, Z\}$.

\subsection{Details of [[8,1,3]] Code}\label{sec:code}
We use the non-CSS code construction method outlined in \cite{shaw} to obtain stabilizer generators for a $[[8,1,3]]$ code.
The code is modified to reduce the weight of the stabilizer generators such that it shows fault-tolerant properties requiring a single ancillary qubit.
Table~\ref{tab:stab_gen} presents the stabilizer generators and the logical $\bar{X}$ and $\bar{Z}$ operators of the code.
\begin{table}[ht]
    \centering
    \begin{tabular}{C|C}
        \hline
        \text{Stabilizer Generators} & \text{Logical Operators}\\ \hline
        Z_0X_1Z_2Z_4  &  \\
        Y_0Y_2Z_3Z_4  &  \\
        Z_0Z_1X_4Z_6  &  \bar{X}=Z_0Z_1X_2Z_5 \\
        X_1Z_3X_5X_6  &  \bar{Z}=Z_0Z_2Z_5Z_6 \\
        Z_3Z_5Z_6X_7  &  \\
        Z_0X_3Z_6Z_7  &  \\
        Z_1X_2X_3Z_4X_6X_7 & \\ \hline
    \end{tabular}
    \caption{Stabilizers and logical operators for [[8,1,3]] code.}
    \label{tab:stab_gen}
\end{table}

We analysed the single ancilla fault-tolerant properties of the code and found that certain propagated errors have similar syndrome values.
For example, while measuring the syndrome for $Z_0X_3Z_6Z_7$ stabilizer, an $X$ error on the ancilla after $X_3$ gate would propagate to $Z_6Z_7$ error on the data qubits. 
Its syndrome is same as $Y_5$ error and after applying correction, we will end up with $Y_5Z_6Z_7$ error which is the logical $\bar{Y}$ error.

We propose to reorder certain stabilizers of [[8,1,3]] code in the following manner:
\begin{align*}
    Z_3Z_5Z_6X_7 &\longrightarrow Z_5Z_3Z_6X_7\\
    Z_0X_3Z_6Z_7 &\longrightarrow Z_7X_3Z_0Z_6\\
    Z_1X_2X_3Z_4X_6X_7 &\longrightarrow Z_1X_3Z_4X_2X_6X_7
\end{align*}
Such a reordering scheme for syndrome measurement leads to distinct syndromes for all possible propagated errors shown in column 2 of Table~\ref{tab:[[8,1,3]]_syndromes}.
The exact syndrome values corresponding to correctable errors and the errors propagated from faults during detection using reordered scheme are given in Section~\ref{sec:lookup_table}.
We notice that all the errors have distinct syndrome values upto a stabilizer.
Thus, we can achieve \textit{fault-tolerance} in the $[[8,1,3]]$ code without requiring extra ancilla qubits.

\begin{table}[htbp]
    \renewcommand{\arraystretch}{1.3}
    \centering
    \begin{tabular}{|C|C|C|}
        \hline
        \rowcolor{yellow} \text{Stabilizers} & \multicolumn{2}{c|}{Errors} \\
        \hline
        Z_0X_1Z_2Z_4 & Z_0P_1 & P_2Z_4 \\ \hline
        Y_0Y_2Z_3Z_4 & Y_0P_2 & P_3Z_4 \\ \hline
        Z_0Z_1X_4Z_6 & Z_0P_1 & P_4Z_6 \\ \hline
        X_1Z_3X_5X_6 & X_1P_3 & P_5X_6 \\ \hline
        Z_5Z_3Z_6X_7 & Z_5P_3 & P_6X_7 \\ \hline
        Z_7X_3Z_0Z_6 & Z_7P_3 & P_0Z_6 \\ \hline
        \multirow{2}{*}{$Z_1X_3Z_4X_2X_6X_7$} & Z_1P_3 & Z_1X_3P_4 \\
        & P_2X_6X_7 & P_6X_7\\
        \hline
    \end{tabular}
    \caption{Propagated Errors where $P\in \{X,Y,Z\}$.}
    \label{tab:[[8,1,3]]_syndromes}
\end{table}

\subsection{Merit of Fault Tolerance}
The [[8,1,3]] code presented is capable of correcting a single qubit error and additionally detect a second error.
The above reordering scheme makes it fault-tolerant without requiring extra ancilla qubits other than the one used while syndrome extraction.
We expect a single fault in the procedure should not lead to a logical error.
The code achieves the same as follows:

\begin{itemize}
     \item All $P \otimes X$ errors are distinguishable by syndromes, upto a stabilizer and a phase, and thus our reordering scheme tackles all hook errors, considering the lookup table decoder.
     \item Any $P \otimes Z$ kind of error leads to a single qubit error in data while also flipping the syndrome value.
     We perform minimum two rounds of syndrome extraction and compare the measurement results.
     Same results ensure no $Z$ error on ancilla while different outcomes require another round of syndrome extraction. When two consecutive syndrome extraction rounds give same outcomes, we  correct the errors accordingly.
     \item Errors of kind $P \otimes Y$ are equal to $(P \otimes X)(I \otimes Z)$. Repeated application of Detector circuit gets rid of $Z$ on ancilla and the syndrome outcome tells the propagated error to be corrected.
\end{itemize}

\subsection{Errors and Syndrome Values}\label{sec:lookup_table}
This section provides the syndrome values corresponding to the single qubit errors correctable by the code and the propagated errors due to faults in error correction procedure.

All single qubit errors are correctable by the [[8,1,3]] code with each error having distinct syndromes given in table~\ref{tab:correctable_errors}.
Errors arising due to faults in detection mainly accumulates through error propagation.
Table~\ref{tab:propagated_error_syndromes} presents syndrome values corresponding to higher weight errors.
We observe that the syndrome corresponding to $Z_1X_3X_4$ is same as $Z_7$ error as $Z_1X_3X_4Z_7$ is an stabilizer.

\begin{table}[ht]
    \centering
    \begin{tabular}{|C|C|C|}
    \hline
    \rowcolor{yellow}\multicolumn{3}{|C|}{\textrm{Single Qubit Errors}}\\ \hline
    X_0: 1110010  &  Y_0: 1010010 & Z_0: 0100000\\ \hline
    X_1: 0010001  &  Y_1: 1011001 & Z_1: 1001000\\ \hline
    X_2: 1100000  &  Y_2: 1000001 & Z_2: 0100001\\ \hline
    X_3: 0101100  &  Y_3: 0101111 & Z_3: 0000011\\ \hline
    X_4: 1100001  &  Y_4: 1110001 & Z_4: 0010000\\ \hline
    X_5: 0000100  &  Y_5: 0001100 & Z_5: 0001000\\ \hline
    X_6: 0010110  &  Y_6: 0011111 & Z_6: 0001001\\ \hline
    X_7: 0000010  &  Y_7: 0000111 & Z_7: 0000101\\
    \hline
    \end{tabular}
    \caption{Syndromes of single-qubit errors.}
    \label{tab:correctable_errors}
\end{table}

\begin{table}[ht]
    \centering
    \begin{tabular}{|C|C|C|C|}
        \hline
        \rowcolor{yellow}\multicolumn{4}{|C|}{\textrm{Errors due to Faults}}\\ \hline
        \rowcolor{yellow}\multicolumn{4}{|C|}{Z_0X_1Z_2Z_4}\\ \hline
        Z_0X_1 & 0110001 & X_2Z_4 & 1110000 \\ \hline
        Z_0Y_1 & 1111001 & Y_2Z_4 & 1010001 \\ \hline
        Z_0Z_1 & 1101000 & Z_2Z_4 & 0110001 \\ \hline
        \rowcolor{yellow}\multicolumn{4}{|C|}{Y_0Y_2Z_3Z_4}\\ \hline
        Y_0X_2 & 0110010 & X_3Z_4 & 0111100 \\ \hline
        Y_0Y_2 & 0010011 & Y_3Z_4 & 0111111 \\ \hline
        Y_0Z_2 & 1110011 & Z_3Z_4 & 0010011 \\ \hline
        \rowcolor{yellow}\multicolumn{4}{|C|}{Z_0Z_1X_4Z_6}\\ \hline
        Z_0X_1 & 0110001 & X_4Z_6 & 1101000 \\ \hline
        Z_0Y_1 & 1111001 & Y_4Z_6 & 1111000 \\ \hline
        Z_0Z_1 & 1101000 & Z_4Z_6 & 0011001 \\ \hline
        \rowcolor{yellow}\multicolumn{4}{|C|}{X_1Z_3X_5X_6}\\ \hline
        X_1X_3 & 0111101 & X_5X_6 & 0010010 \\ \hline
        X_1Y_3 & 0111110 & Y_5X_6 & 0011010 \\ \hline
        X_1Z_3 & 0010010 & Z_5X_6 & 0011110 \\ \hline
        \rowcolor{yellow}\multicolumn{4}{|C|}{Z_5Z_3Z_6X_7}\\ \hline
        Z_5X_3 & 0100100 & X_6X_7 & 0010100 \\ \hline
        Z_5Y_3 & 0100111 & Y_6X_7 & 0011101 \\ \hline
        Z_5Z_3 & 0001011 & Z_6X_7 & 0001011 \\ \hline
        \rowcolor{yellow}\multicolumn{4}{|C|}{Z_7X_3Z_0Z_6}\\ \hline
        Z_7X_3 & 0101001 & X_0Z_6 & 1111011 \\ \hline
        Z_7Y_3 & 0101010 & Y_0Z_6 & 1011011 \\ \hline
        Z_7Z_3 & 0000110 & Z_0Z_6 & 0101001 \\ \hline
        \rowcolor{yellow}\multicolumn{4}{|C|}{Z_1X_3Z_4X_2X_6X_7}\\ \hline
        Z_1X_3 & 1100100 & X_6X_7 & 0010100 \\ \hline
        Z_1Y_3 & 1100111 & Y_6X_7 & 0011101 \\ \hline
        Z_1Z_3 & 1001011 & Z_6X_7 & 0001011 \\ \hline
        Z_1X_3X_4 & 0000101 & X_2X_6X_7 & 1110100 \\ \hline
        Z_1X_3Y_4 & 0010101 & Y_2X_6X_7 & 1010101 \\ \hline
        Z_1X_3Z_4 & 1110100 & Z_2X_6X_7 & 0110101 \\ \hline
    \end{tabular}
    \caption{Syndromes of propagated errors.}
    \label{tab:propagated_error_syndromes}
\end{table}

\subsection{Quantum Circuits}\label{sec:circuits}
This section presents the quantum circuits used in simulating the $[[8,1,3]]$ code with the noise models as described in section~\ref{sec:noise_models}.

\begin{figure*}[htbp]
    \centering
    \includegraphics[width=\linewidth]{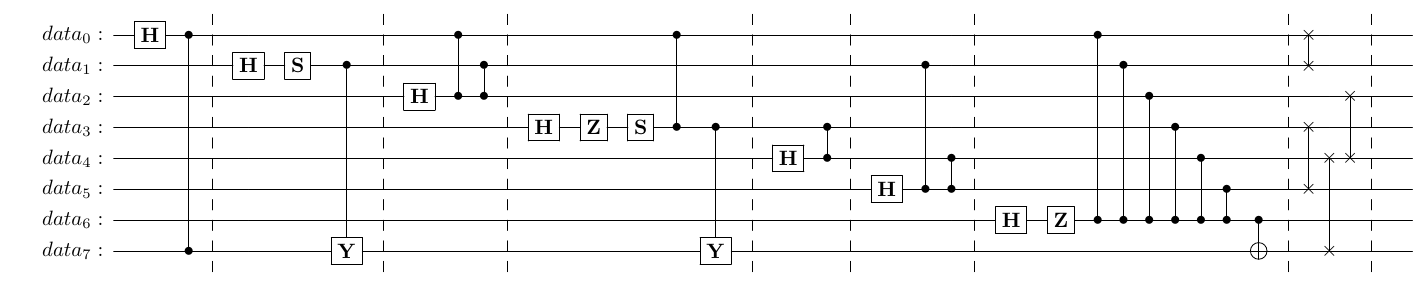}
    \caption{Encoder Circuit for $[[8,1,3]]$ code. The last $k=1$ qubit is initialized in $\ket{\psi_i}$ state.}
    \label{fig:encoder_circuit}
\end{figure*}

We use the methods presented in Section~\ref{sec:encoder} to create an efficient unitary Encoder for the chosen stabilizers.
The Encoder circuit given in Fig.~\ref{fig:encoder_circuit} utilizes the enhanced standard form mechanism for non-CSS codes without any non-essential gates.
We notice the last step involves certain \textit{SWAP} gates whose purpose is to map relabeled qubits to the original order of qubits.

\begin{figure*}[htbp]
    \centering
    \includegraphics[width=\linewidth]{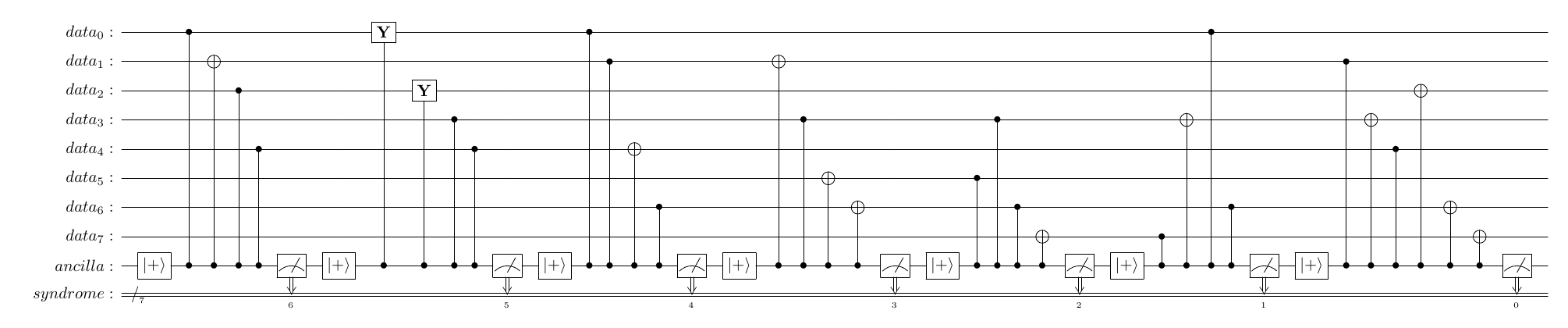}
    \caption{Detector circuit for syndrome measurements, incorporating reordering scheme for [[8,1,3]] code. All measurements are in $X$ basis followed by resetting the qubit to $\ket{+}$ state.}
    \label{fig:detector_circuit}
\end{figure*}

The syndrome measurement circuit, here referred to as \textit{Detector}, is built following the reordering scheme presented in 
Section 
\ref{sec:code}.
We utilize fast reset of qubits to use a single ancilla for syndrome measurements.
It enhances the simulation speed exponentially as fewer space resources are required to store quantum states in classical systems.
The complete circuit is shown in Fig. ~\ref{fig:detector_circuit}.

The corrector circuit is a classical-quantum hybrid circuit that applies certain fixed operations on the data qubits given the syndrome values.
We note that we are not using a standard lookup table decoder that contains the most likely error configuration for every possible syndrome value.
Instead, our lookup table only consists of syndrome and error pairs as given in Section~\ref{sec:lookup_table}.

One can test the efficiency of the code under different decoders for correcting errors.
Details of the error rate calculation and simulation runs are given in Section~\ref{sec:simulation}.
\section{Noise Models}\label{sec:noise_models}
To study the performance of the proposed $[[8,1,3]]$ code, we test it on two noise models: the standard depolarising noise model and the anisotropic noise model.

\subsection{Standard Depolarising Noise Model}\label{sec:std_dep_noise}
It is the most applicable noise model to test physical systems, in which each gate is followed by a symmetric depolarization with probability $p$.
Thus, with probability $p$, the state is replaced by $I/2$ and is left the same by probability $1-p$.
The simulation model of same is as follows:
    \begin{enumerate}
        \item With probability $p_s$, each single qubit gate is followed by a single qubit Pauli error drawn uniformly and independently from $\{X,Y,Z\}$. Thus either of the three error occurs with probability $p_s/3$.
        \item With probability $p_t$, each two qubit gate is followed by a two qubit Pauli error drawn uniformly and independently from $\{I,X,Y,Z\} \otimes \{I,X,Y,Z\} \setminus \{I \otimes I\}$. Each of 15 errors occurs with probability $p_t/15$.
        \item With probability $p_{meas}$, each single qubit measurement has its outcome flipped.
        \item With probability $p_{prep}$, each preparation of $\ket{0}$ is flipped to $\ket{1}$ and each preparation of $\ket{+}$ is flipped to $\ket{-}$.
    \end{enumerate}

\subsection{Anisotropic Noise Model}\label{sec:anisotropic_noise}
A suitable noise model that mimics over or under-rotation of gates in ion-trap qubits is the anisotropic noise model.
It considers noise due to gate coupling and thus two qubit errors are only the ones aligned with gates \cite{bare_ancilla}.
Anisotropic noise model is simulated as follows:
\begin{enumerate}
    \item Single qubit error, readout error, and preparation error are applied with probability $p_s$, $p_{meas}$, and $p_{prep}$ respectively as defined above for Standard Depolarising Noise model.
    \item With probability $p_t$, every two qubit $CP$ (controlled-$P$) gate is first followed by $Z \otimes P$ error. Each of the two qubits involved are then acted upon by a single qubit error with probability $p_s$.
\end{enumerate}
For our purpose, we have taken all probabilities to be same, i.e., $p_s = p_t = p_{meas} = p_{prep}$, in each of the noise model. A further robust noise model may consider single qubit errors on idle locations too.

\section{Simulation Scheme}\label{sec:simulation}
We designed the complete error correction simulator using qiskit following the scheme described in Fig.~\ref{fig:error_correction_picture}\cite{Qiskit}.
We consider $\ket{\psi_i} = \ket{0}$ and initialize state $\ket{0}^{\otimes 8}$, we apply the Encoder circuit which creates the logical $\ket{\bar{0}}$ state for the given code.
The Detector circuit is then applied twice.
The classical outputs of the two Detector circuits are compared, if they are same we proceed with Corrector, else we apply Detector circuit again and feed its output to the corrector.
Ideally, we should keep applying the Detector until two consecutive outcomes are identical.
Due to resource constraints, we limit our simulation to a maximum of three Detector applications.
The error-corrected state is fed into a circuit which is the adjoint of the Encoder, and the data qubits are measured.
For our purpose, the Encoder, the Corrector, and the adjoint of the Encoder are kept ideal, i.e., noise-free. The Detector circuit can observe errors which follow from the noise models as described in Sections~\ref{sec:std_dep_noise} and ~\ref{sec:anisotropic_noise}.

\subsection{Error Rate Calculation}\label{sec:error_rate_calculation}
We analyze three metrics for the code on each noise model.
\begin{enumerate}
\item Fidelity: Probability of the decoded state, $\ket{\psi_d}$ being same as the encoded state, $\ket{\psi_i}$.
For practical quantum computation, we require as high fidelity as possible.

\item Logical error rate: Probability of the decoded state to remain in the codespace but different from the encoded state i.e., $\ket{\psi_i} \ne \ket{\psi_d} \in \mathcal{C}$.
A logical error is an unwanted operation which will always go undetected, altering the state of interest and hence giving wrong computational results.
The lower the logical error rate, the better the code.

\item Total error rate: Probability of the decoded state being different from the encoded state.
It encompasses logical and other errors that brings the state out of codespace.
We can trivially say total error rate = 1-fidelity.
\end{enumerate}
There are three possible measurement outcomes of data qubits.
If the first $n-k$ qubits are in state $\ket{0}$, the decoded state, $\ket{\psi_d} \in \mathcal{C}$.
Otherwise, the error-corrected state is outside the codespace.
In reverse order of qubit measurements, $\ket{0}\ket{0}^{\otimes 7}$ contributes to fidelity, while $\ket{1}\ket{0}^{\otimes 7}$ contributes to logical error.
The complete simulation requires the following steps:
\begin{enumerate}
    \item For a given value of physical error rate, $p$, run the simulation circuit $N$ (atleast $100\times p$) times and note down the outcomes.
    \item Define logical error rate = (no. of logical errors / no. of times the circuit was run) and fidelity = (no. of times same as initial state occurs / no. of times the circuit was run).
    \item Repeat step 1 and 2 for $m$ rounds to obtain a distribution of error rate values at a given $p$. The mean of $m$ rounds along with minimum and maximum value is plotted against the physical error rate.
    \item Repeat step 1, 2, and 3 for several different values of $p$.
\end{enumerate}
Once we have obtained the metrics for a range of values of $p$, we fit a polynomial curve to judge the pseudo-thresholds.
The error correcting scheme is expected to reduce any linear dependence of logical or total error rates on the physical error rate.
It is encouraged to fit a curve of form $a_0x^2 + a_1x^3 + \dots + a_{n-2}x^n$ where $a_0$ is the leading order coefficient.
Since we initialize the state in logical $\ket{\bar{0}}$ state, the effect of logical $Z$ operator will not be noticed and only logical $X$ and logical $Y$ will cause an observable logical error.
Thus we compare the obtained error rates to $\frac{2}{3}p$ instead of $p$.

The pseudo-threshold is the value at which the error rate after error correction crosses the unencoded error rate ($\frac{2}{3}p$).
It denotes the value of physical error rate below which the code in test would be more suitable for performing quantum computation than without any encoding.
Comparing two codes, a code with higher pseudo-threshold will be preferable.
The leading coefficient defined above is another metric to compare codes. The lower the coefficient, the better performing the code.

\subsection{Modified Simulation}\label{sec:modified_simulation}
The above method is more realistic to physical devices.
Nonetheless, it does not account for all correctable errors.
QEC simulations often involves a noise-free round of projecting the error-corrected state to the codespace and comparing it to the encoded state.
This involves additional noise-free error detection and correction circuit before $\text{(Encoder)}^{\dagger}$ operation.

The modified approach leads to a state containing only those errors not correctable by the code.
This leads to an increase in fidelity and an increase in logical error rate as errors of comprising a logical operator times a correctable error leads to logical error in measurement output.

\section{Discussion of the results}\label{sec:results}
This section presents results obtained from simulation runs and the properties of the code obtained from analysing results.
It also includes the authors' perspective on the results and comparison.
We call the simulation run without the noise-free projection step described in Section~\ref{sec:error_rate_calculation} as the \textit{Practical Method}.
And the simulation run with the noise-free detection-correction step described in Section~\ref{sec:modified_simulation} as the \textit{Modified Method}.
We obtain \textit{pseudo-threshold} for the code in both the noise models.
We also obtain the \textit{leading order terms} from the best-fit curve.

\begin{figure}[htb]
    \centering
    % \centerline{
        \subfigure[Logical error rates vs Physical error rates.]{
            \includegraphics[width=\linewidth]{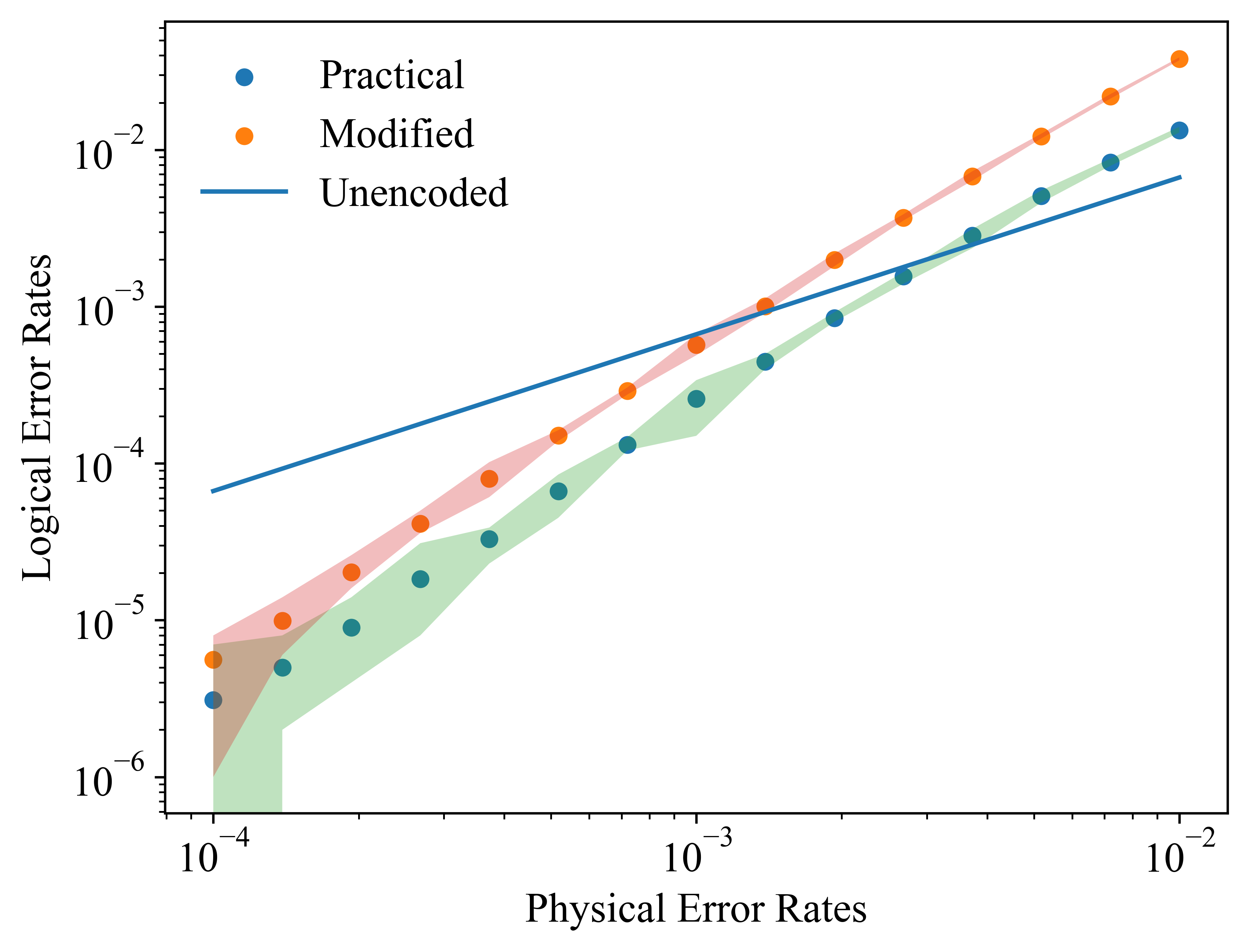}
            \label{fig:std_dep_logical}
        }
        \subfigure[Total error rates vs Physical error rates.]{    
            \includegraphics[width=\linewidth]{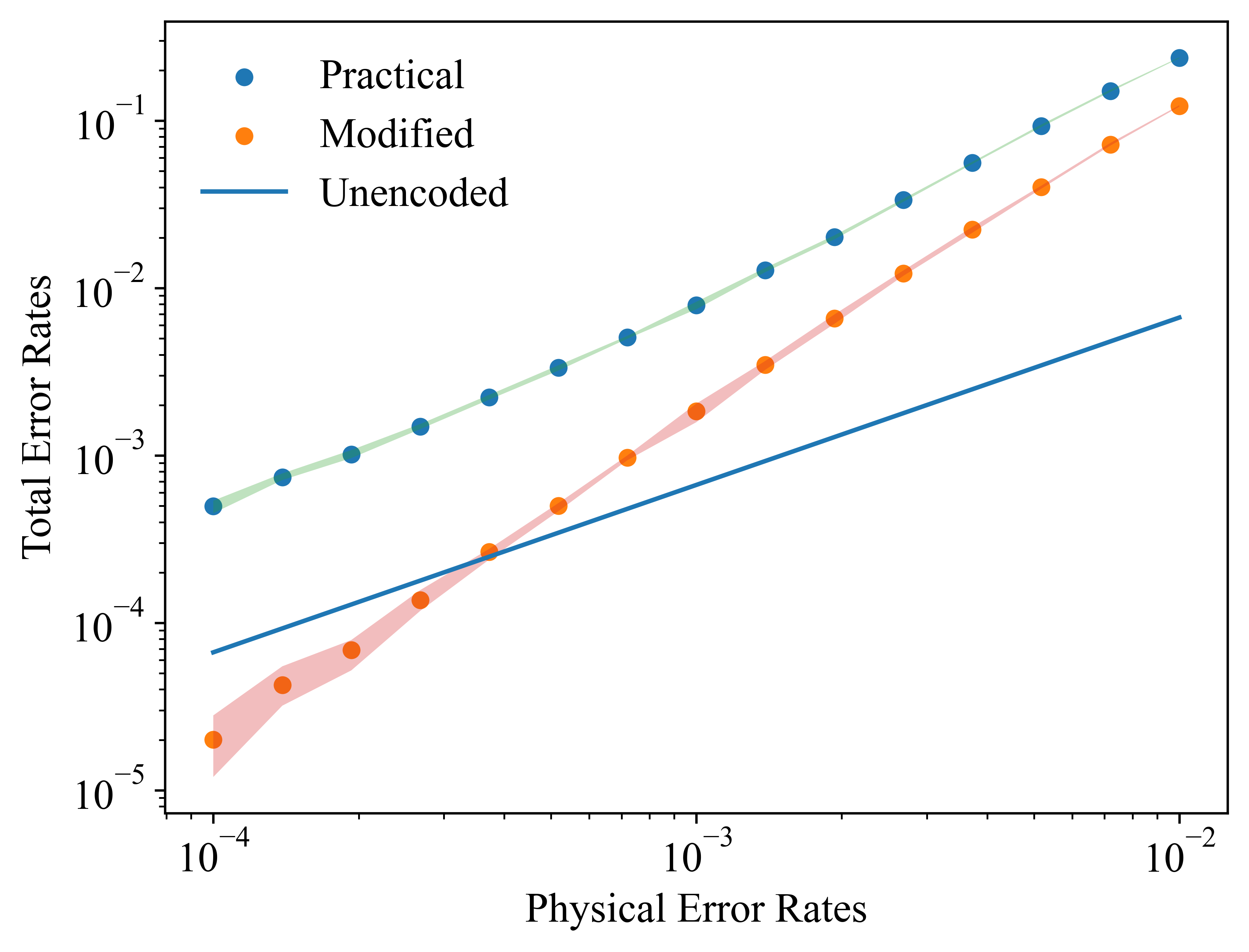}
            \label{fig:std_dep_total}
        }
    % }
    \caption{Error rates for standard depolarising noise model. Best fit curve is shown for both methods for logical error rates and for modified method in total error rate.}
    \label{fig:std_dep_plots}
\end{figure}

% Comparison of Error Rate metrics for our realistic simulation run and the modified simulation run. The solid curve represents the best fit curve and the distribution shows the minimum and maximum value of respective error rate obtained in $m=15$ and $m=10$ rounds for practical and modified runs.
% \subsection{Standard depolarising Noise}
We first analyze the performance of the code in a standard depolarising noise model.
Fig.~\ref{fig:std_dep_plots} presents the descent of logical and total error rates compared among the two simulation methods.
We obtain leading order terms for the practical and modified simulation runs with respective logical error threshold values of 0.3\% and 0.12\%.
The practical method could not provide a pseudo-threshold for total error rates, thus showing a linear relation to unencoded error rates.
The modified approach nonetheless provides a threshold of 0.034\%.

\begin{table}[ht]
    \centering
    \begin{tabular}{|c|c|c|c|}
        \hline
        \rowcolor{yellow}\multicolumn{2}{|c|}{Error Rates} & Pseudo-threshold & Leading Order \\ \hline
        \multirow{2}{*}{Logical Error Rates} &
        Practical & 0.003145 & 270 \\
        & Modified & 0.001212 & 550 \\ \hline
        \multirow{2}{*}{Total Error Rates} &
        Practical & \--  & \-- \\
        & Modified & 0.0003426  & 2016 \\ \hline
    \end{tabular}
    \caption{Metrics for standard depolarising noise model.}
    \label{tab:std_dep_results}
\end{table}

\begin{figure}[htbp]
    \centering
    % \centerline{
        \subfigure[Logical error rates vs Physical error rates.]{
            \includegraphics[width=\linewidth]{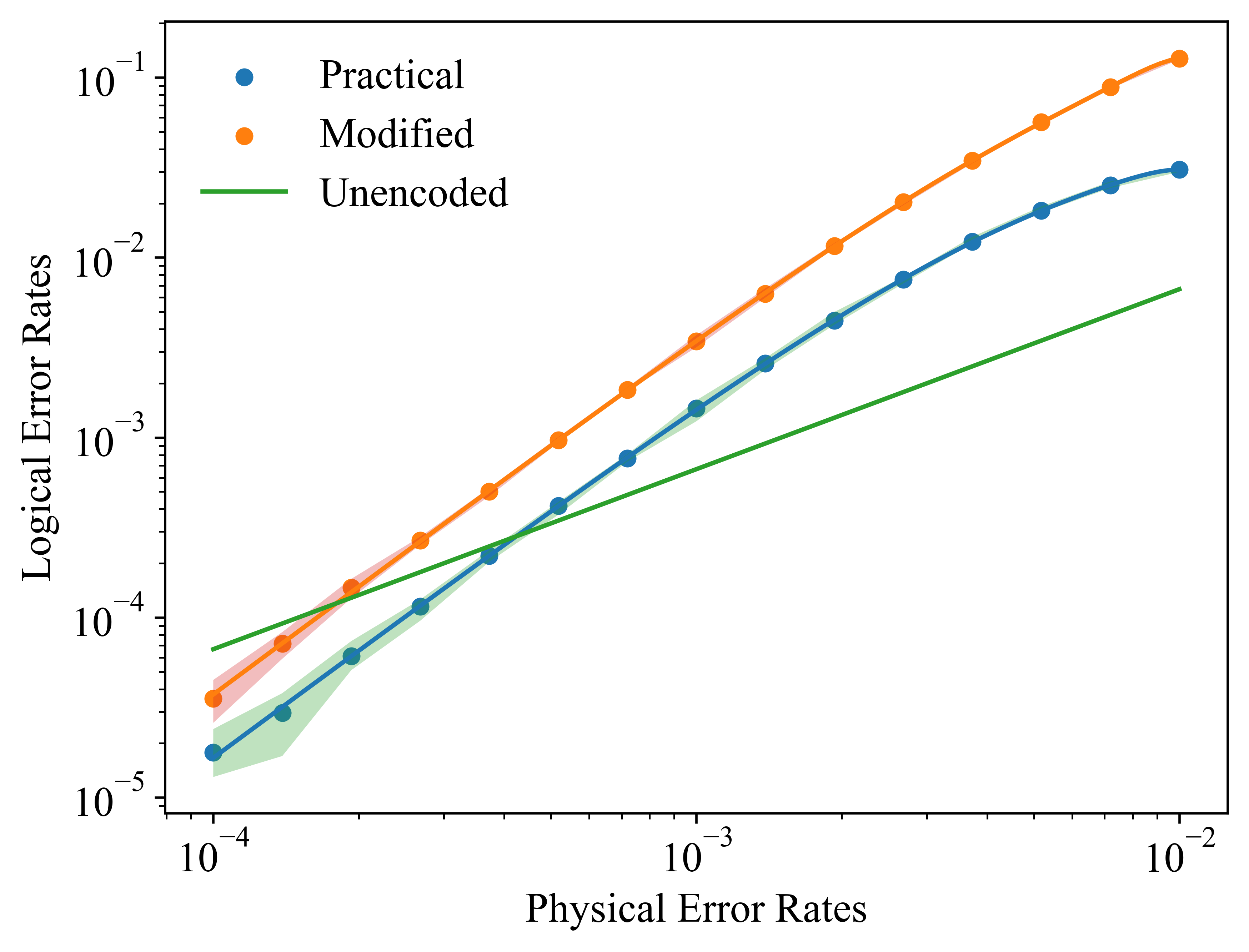}
            \label{fig:anisotropic_logical}
        }
        \subfigure[Total error rates vs Physical error rates.]{    
            \includegraphics[width=\linewidth]{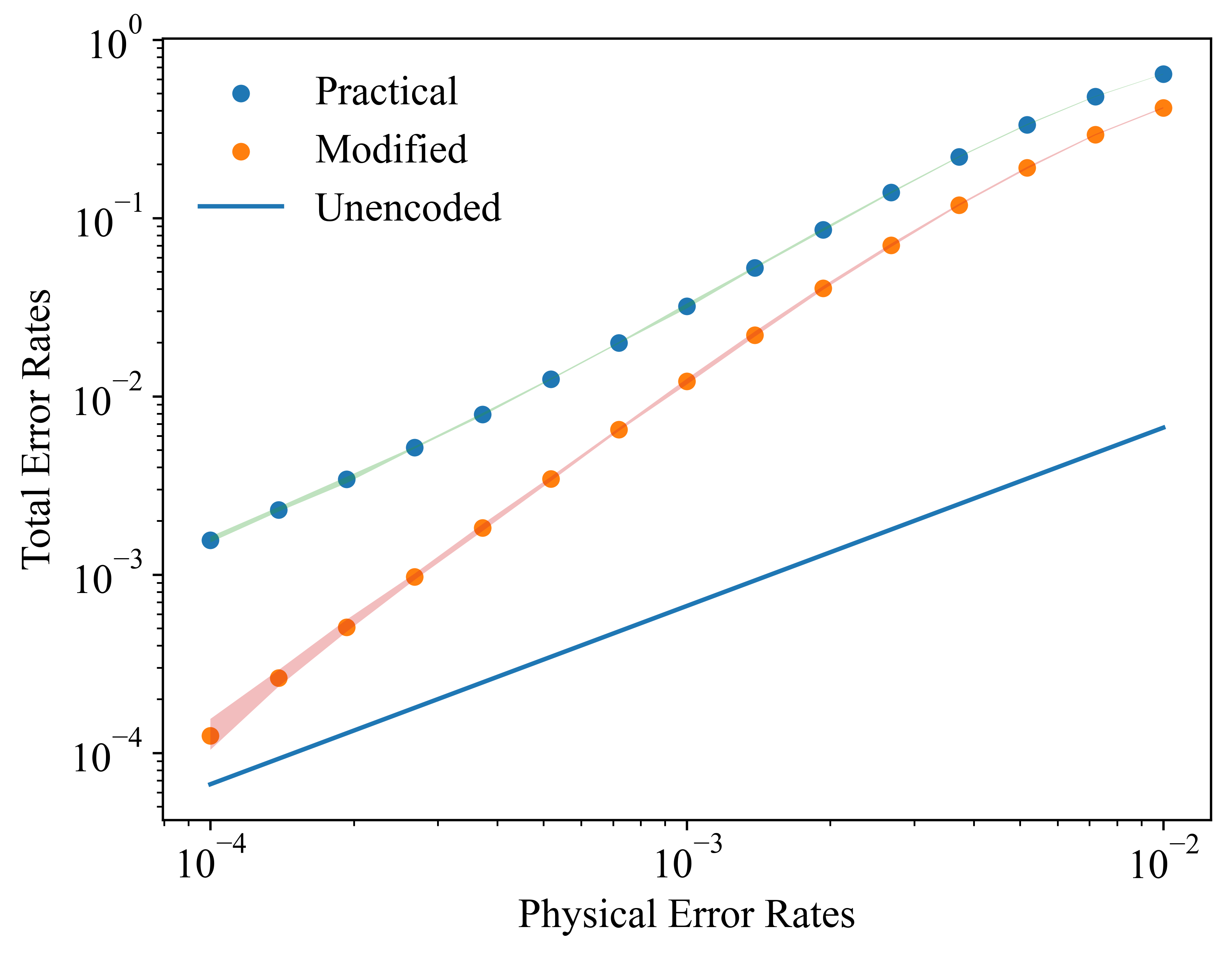}
            \label{fig:anisotropic_total}
        }
    % }
    \caption{Error rates for anisotropic noise model. Best fit curve is shown for both methods for logical error rates and for modified method in total error rate.}
    \label{fig:anisotropic_plots}
\end{figure}

% \subsection{Anisotropic Noise}
For the anisotropic noise model, the descent of logical and total error rates is presented in Fig.~\ref{fig:anisotropic_plots}.
The threshold values for the practical and modified methods are found to be 0.042\% and 0.018\% respectively for logical error rates.
The practical method fails again with total error rates, but the modified method provides a threshold of $4.95\times 10^{-3}\%$ on extrapolating the best-fit curve.

\begin{table}[ht]
    \centering
    \begin{tabular}{|c|c|c|c|}
        \hline
        \rowcolor{yellow}\multicolumn{2}{|c|}{Error Rates} & pseudo-threshold & Leading Order \\ \hline
        \multirow{2}{*}{Logical Error Rates} &
        Practical & 0.0004239 & 1689 \\
        & Modified & 0.0001796 & 3767 \\ \hline
        \multirow{2}{*}{Total Error Rates} &
        Practical & \--  & \-- \\
        & Modified & $4.95 \times 10^{-5}$  & 13540 \\ \hline
    \end{tabular}
    \caption{Metrics for anisotropic noise model.}
    \label{tab:anisotropic_results}
\end{table}

\begin{figure}[htbp]
    \centering
    % \centerline{
        \subfigure[Leading orders of logical error rates.]{
            \includegraphics[width=\linewidth]{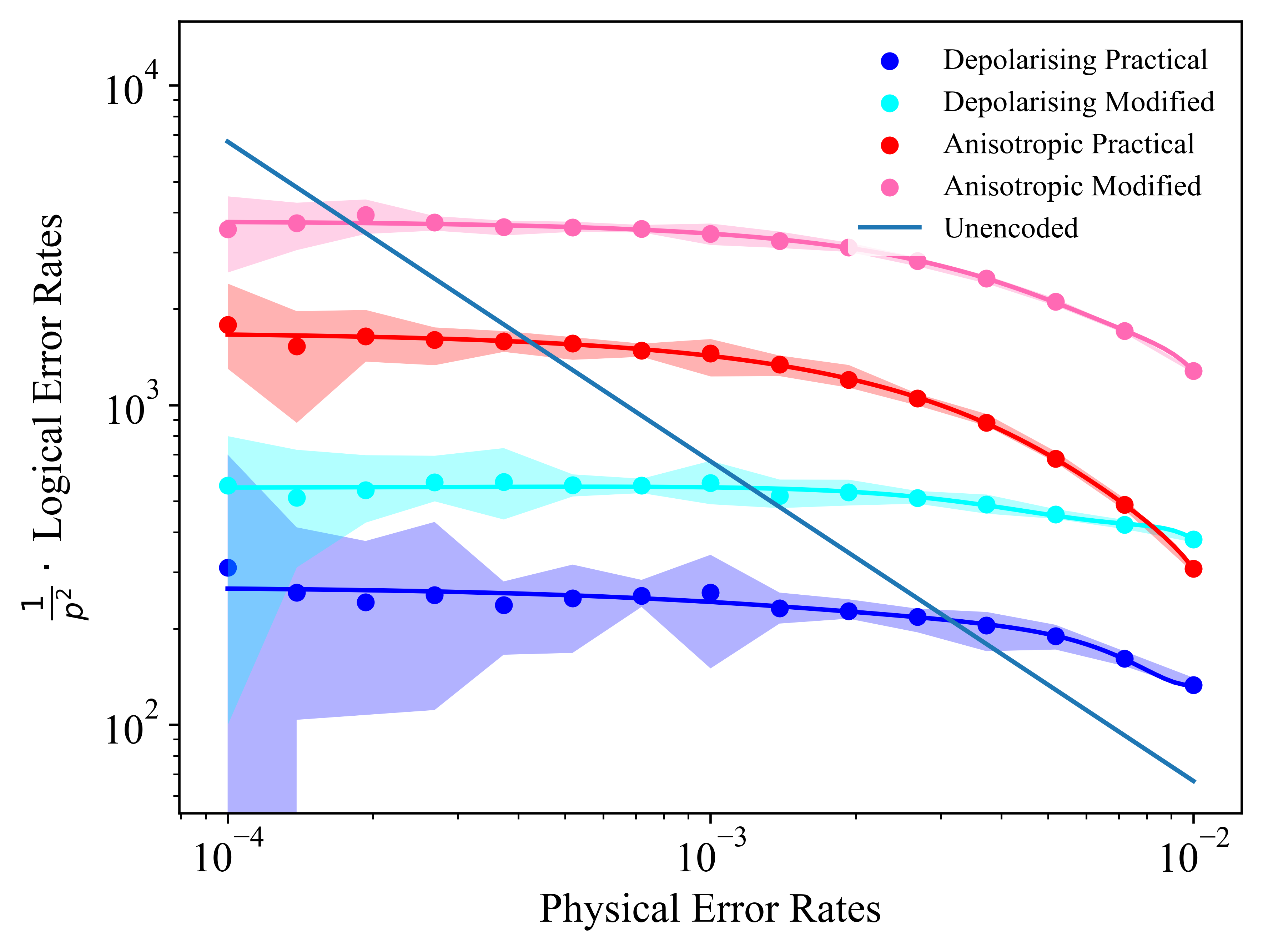}
            \label{fig:logical_leading_order}
        }
        \subfigure[Leading orders of total error rates.]{    
            \includegraphics[width=\linewidth]{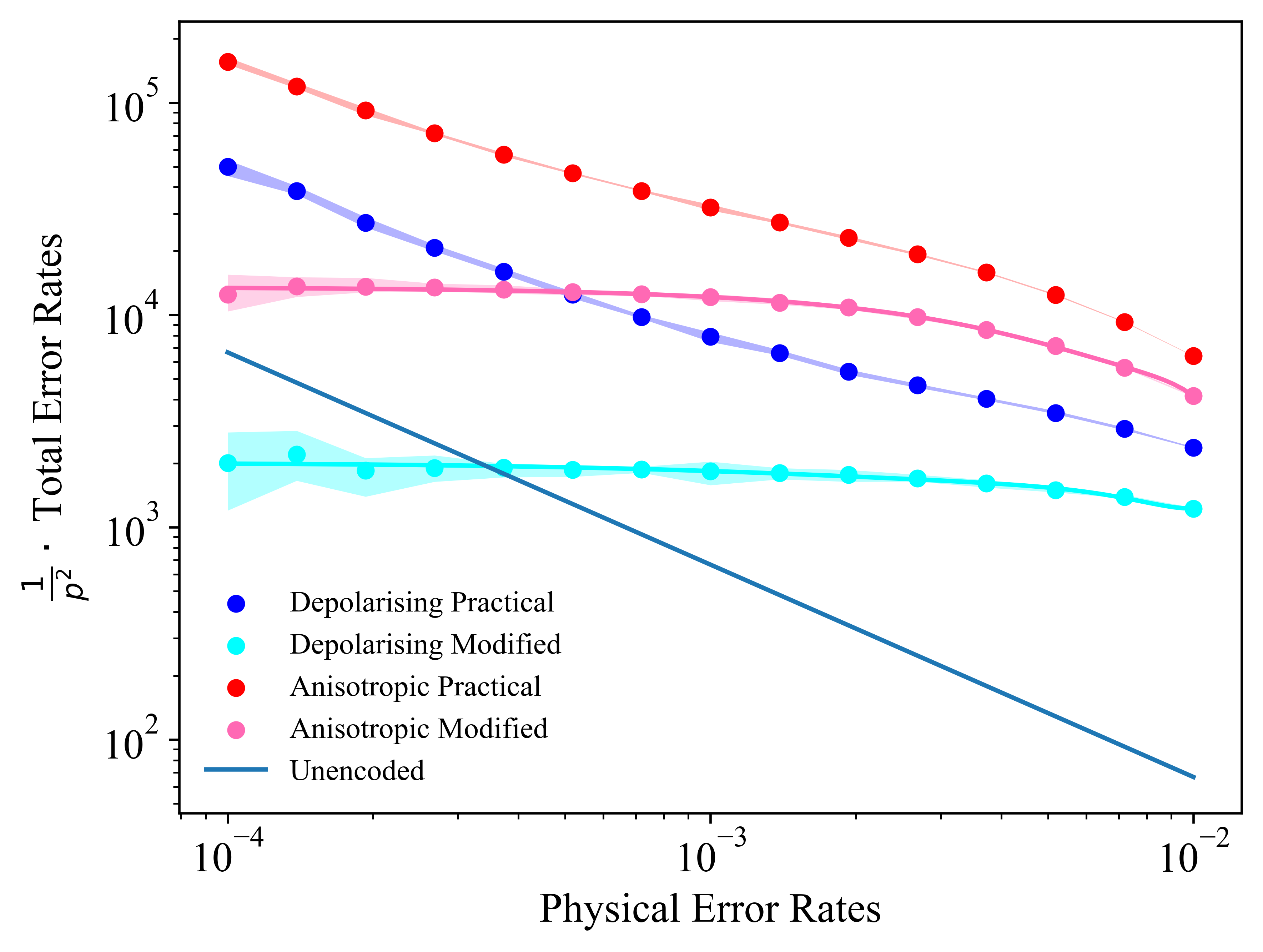}
            \label{fig:total_leading_order}
        }
    % }
    \caption{Leading order of error rates with both simulation methods and noise models. The distribution shows the minimum to maximum range of error rates around the plotted mean value.}
    \label{fig:leading_orders}
\end{figure}

Fig.~\ref{fig:leading_orders} presents the error rates divided by $p^2$ to reveal the leading order of the best fit curve.
Logical error rates crosses the unencoded error rates for both noise models and for both simulation methods.
Total error rates shows the dependence of practical method results on physical error rates while the modified method leads to a distinctive convergence for the leading coefficient of $p^2$.

Evidently, the code performance is superior in the standard depolarising model than the anisotropic model.
This can be attributed to the error correction capabilities of the code as well the noise in the model.
The anisotropic noise has a two qubit error followed by single qubit errors increasing the chance of $Z$ error on the syndrome which can only be corrected by consecutive rounds of error detection.
The constraint of three rounds allow many $Z$ errors to go uncorrected.
On the other hand, the code is well structured to identify any $P \otimes X$ error, getting rid of hook errors and increasing the pseudo-threshold.

While the practical method of simulation is more appropriate to realistic settings, the modified method is more suitable to characterize code performance and provide benchmarks.
Modified method leaves the code with only uncorrectable errors, either logical errors or higher weight errors, thus it puts a bar on the efficiency of the code.
We further observe that the leading order of $p^2$ is also an appropriate metric to compare performance of a code.
The lower the leading order coefficient, the higher is the pseudo-threshold value.

\section{Conclusions}
We used the $[[8,1,3]]$ non-CSS code and showed Fault-tolerance using the modified Bare ancillary method of \cite{bare_ancilla}. 
We obtained pseudo-thresholds of the code under the depolarising and anisotropic noise models.
Our fault tolerant scheme is much suitable to standard depolarising noise and provides higher pseudo-threshold values than widely studied fault tolerant techniques implemented on various codes \cite{Chamberland_2018}.
We compared the practical error correction procedure to the traditional simulation run and compared the merits of both, presenting the requirement of noise-free projection to encoded state for reliable simulation results.

A promising future direction would be studying the code performance under various noise models and to generalize the single qubit fault tolerance method to other codes.

\section*{Acknowledgment}
The authors thank Harsh Gupta for implementing methods given in \cite{shaw} and providing us with the stabilizers of a non-CSS [[8,1,3]] code.
We thank Krishnakumar Sabapathy for encouraging us to study Bare ancillary fault tolerance method \cite{bare_ancilla} and providing fruitful comments in the manuscript preparation.

% \enlargethispage{-4cm}
% \pagebreak
% \IEEEtriggeratref{8}
\bibliographystyle{IEEEtran}
\bibliography{bibliofile}

\begin{thebibliography}{10}
\providecommand{\url}[1]{#1}
\csname url@samestyle\endcsname
\providecommand{\newblock}{\relax}
\providecommand{\bibinfo}[2]{#2}
\providecommand{\BIBentrySTDinterwordspacing}{\spaceskip=0pt\relax}
\providecommand{\BIBentryALTinterwordstretchfactor}{4}
\providecommand{\BIBentryALTinterwordspacing}{\spaceskip=\fontdimen2\font plus
\BIBentryALTinterwordstretchfactor\fontdimen3\font minus \fontdimen4\font\relax}
\providecommand{\BIBforeignlanguage}[2]{{%
\expandafter\ifx\csname l@#1\endcsname\relax
\typeout{** WARNING: IEEEtran.bst: No hyphenation pattern has been}%
\typeout{** loaded for the language `#1'. Using the pattern for}%
\typeout{** the default language instead.}%
\else
\language=\csname l@#1\endcsname
\fi
#2}}
\providecommand{\BIBdecl}{\relax}
\BIBdecl

\bibitem{CSS_shor}
\BIBentryALTinterwordspacing
A.~R. Calderbank and P.~W. Shor, ``Good quantum error-correcting codes exist,'' \emph{Phys. Rev. A}, vol.~54, pp. 1098--1105, Aug 1996. [Online]. Available: \url{https://link.aps.org/doi/10.1103/PhysRevA.54.1098}
\BIBentrySTDinterwordspacing

\bibitem{CSS_Steane}
\BIBentryALTinterwordspacing
A.~M. Steane, ``Simple quantum error-correcting codes,'' \emph{Phys. Rev. A}, vol.~54, pp. 4741--4751, Dec 1996. [Online]. Available: \url{https://link.aps.org/doi/10.1103/PhysRevA.54.4741}
\BIBentrySTDinterwordspacing

\bibitem{aharonov}
\BIBentryALTinterwordspacing
D.~Aharonov and M.~Ben-Or, ``Fault-tolerant quantum computation with constant error rate,'' \emph{SIAM Journal on Computing}, vol.~38, no.~4, pp. 1207--1282, 2008. [Online]. Available: \url{https://doi.org/10.1137/S0097539799359385}
\BIBentrySTDinterwordspacing

\bibitem{nielsen_chuang_2010}
M.~A. Nielsen and I.~L. Chuang, \emph{Quantum Computation and Quantum Information: 10th Anniversary Edition}.\hskip 1em plus 0.5em minus 0.4em\relax Cambridge University Press, 2010.

\bibitem{ft_shor}
\BIBentryALTinterwordspacing
P.~Shor, ``Fault-tolerant quantum computation,'' in \emph{2013 IEEE 54th Annual Symposium on Foundations of Computer Science}.\hskip 1em plus 0.5em minus 0.4em\relax Los Alamitos, CA, USA: IEEE Computer Society, oct 1996. [Online]. Available: \url{https://doi.ieeecomputersociety.org/10.1109/SFCS.1996.548464}
\BIBentrySTDinterwordspacing

\bibitem{ft_steane}
\BIBentryALTinterwordspacing
A.~M. Steane, ``Active stabilization, quantum computation, and quantum state synthesis,'' \emph{Phys. Rev. Lett.}, vol.~78, pp. 2252--2255, Mar 1997. [Online]. Available: \url{https://link.aps.org/doi/10.1103/PhysRevLett.78.2252}
\BIBentrySTDinterwordspacing

\bibitem{ft_2qubits}
R.~Chao and B.~W. Reichardt, ``Quantum error correction with only two extra qubits,'' \emph{Phys. Rev. Lett.}, vol. 121, p. 050502, 8 2018.

\bibitem{bare_ancilla}
M.~Li, M.~Guti'errez, S.~E. David, A.~Hernandez, and K.~R. Brown, ``Fault tolerance with bare ancillary qubits for a [[7,1,3]] code,'' \emph{Phys. Rev. A}, vol.~96, p. 032341, 9 2017.

\bibitem{shaw}
S.~Shaw, H.~Gupta, S.~M. Shah, and A.~Raina, ``Construction of non-css quantum codes using measurements on cluster states,'' 2023.

\bibitem{Qiskit}
{Qiskit contributors}, ``Qiskit: An open-source framework for quantum computing,'' 2023.

\bibitem{shor_code}
P.~W. Shor, ``Scheme for reducing decoherence in quantum computer memory,'' \emph{Phys. Rev. A}, vol.~52, pp. R2493--R2496, 10 1995.

\bibitem{steane_code}
\BIBentryALTinterwordspacing
A.~M. Steane, ``Error correcting codes in quantum theory,'' \emph{Phys. Rev. Lett.}, vol.~77, pp. 793--797, Jul 1996. [Online]. Available: \url{https://link.aps.org/doi/10.1103/PhysRevLett.77.793}
\BIBentrySTDinterwordspacing

\bibitem{gottesman_thesis}
D.~Gottesman, ``Stabilizer codes and quantum error correction,'' Ph.D. dissertation, California Institute of Technology, California, 1997.

\bibitem{gaitan}
F.~Gaitan, \emph{Quantum Error Correction and Fault Tolerant Quantum Computing}.\hskip 1em plus 0.5em minus 0.4em\relax CRC Press, 2013.

\bibitem{gottesman_encoder}
R.~Cleve and D.~Gottesman, ``Efficient computations of encodings for quantum error correction,'' \emph{Phys. Rev. A}, vol.~56, pp. 76--82, 7 1997.

\bibitem{grassl_encoder}
M.~Grassl, ``Variations on encoding circuits for stabilizer quantum codes,'' in \emph{Coding and Cryptology}.\hskip 1em plus 0.5em minus 0.4em\relax Springer Berlin Heidelberg, 2011, pp. 142--158.

\bibitem{Chamberland_2018}
\BIBentryALTinterwordspacing
C.~Chamberland and M.~E. Beverland, ``Flag fault-tolerant error correction with arbitrary distance codes,'' \emph{Quantum}, vol.~2, p.~53, Feb. 2018. [Online]. Available: \url{http://dx.doi.org/10.22331/q-2018-02-08-53}
\BIBentrySTDinterwordspacing

\end{thebibliography}
\end{document}